\documentclass[12pt, preprint]{aastex}

\newcommand{\spi}{{\em Spitzer}}
\newcommand{\msx}{{\em MSX}}
\newcommand{\mum}{{\micron}}
\newcommand{\ms}{MSX SMC 029}
\newcommand{\mo}{\ifmmode{\rm M_{\odot}}\else{M$_{\odot}$}\fi}
\newcommand{\ben}{\ifmmode{\rm C_6H_6}\else{C$_6$H$_6$}\fi}
\newcommand{\acet}{\ifmmode{\rm C_2H_2}\else{C$_2$H$_2$}\fi}
\newcommand{\diac}{\ifmmode{\rm C_4H_2}\else{C$_4$H$_2$}\fi}
\newcommand{\cyac}{\ifmmode{\rm HC_3N}\else{HC$_3$N}\fi}
\newcommand{\lsun}{\ifmmode{L_\sun}\else{$L_\sun$}\fi}

\begin{document}

\title{A Post-AGB Star in the Small Magellanic Cloud Observed with 
 the {\em SPITZER} Infrared Spectrograph}

\author{Kathleen E. Kraemer\altaffilmark{1}, G. C. Sloan\altaffilmark{2}, 
J. Bernard-Salas\altaffilmark{2}, Stephan D. Price\altaffilmark{1}, Michael 
P. Egan\altaffilmark{1}, P. R. Wood\altaffilmark{3}}

\begin{abstract}
We have observed an evolved star with a rare combination of spectral features,
 \ms, in the Small Magellanic Cloud (SMC) 
using the low-resolution modules of the Infrared Spectrograph on the 
{\em Spitzer Space Telescope}. A cool dust continuum dominates the spectrum
of \ms. The spectrum also shows both emission from polycyclic aromatic
hydrocarbons (PAHs) and absorption at 13.7 \mum\ from \acet, a 
juxtaposition seen in only two other sources, AFGL 2688 and IRAS 13416$-$6243,
both post-asymptotic giant branch (AGB) objects. As in these sources, the PAH 
spectrum has the unusual trait that the peak emission in the 7--9 \mum\
complex lies beyond 8.0 \mum. In addition, the 8.6 \mum\ feature has an
intensity as strong as the C--C modes which normally peak between 7.7 and 7.9
\mum. The relative flux of the feature at 11.3 \mum\ to that at 8 \mum\ 
suggests that the PAHs in \ms\ either have a low ionization fraction or are
largely unprocessed. The 13--16 \mum\ wavelength region shows strong
absorption features similar to those observed in the post-AGB objects
AFGL 618 and SMP LMC 11. This broad absorption may arise from the
same molecules which have been identified in those sources: \acet, \diac, 
\cyac, and \ben. The similarities between \ms, AFGL 2688, and AFGL 618 lead
us to conclude that \ms\ has evolved off the AGB in only the past few
hundred years, making it the third post-AGB object identified in the SMC.

\end{abstract}


\altaffiltext{1}{Air Force Research Laboratory, Space Vehicles 
 Directorate, 29 Randolph Rd., Hanscom AFB, MA 01731; 
kathleen.kraemer@hanscom.af.mil, steve.price@hanscom.af.mil, michael.egan@osd.mil}

\altaffiltext{2}{Department of Astronomy, Cornell University, 108 Space Sciences Building, Ithaca, NY 14853; sloan@isc.astro.cornell.edu, jbs@isc.astro.cornell.edu}

\altaffiltext{3}{Research School of Astronomy \& Astrophysics, Mount Stromlo 
Observatory, Weston Creek, ACT 2611, Australia; wood@mso.anu.edu.au}

\section{Introduction \label{sec.intro}}

As a star ascends the asymptotic giant branch (AGB), its outer
atmosphere expands and pulsates, pushing gas away from the star
where it can cool and condense into dust grains.  The resulting
circumstellar dust shell hides the star in the optical and
emits strongly in the infrared (IR).  Eventually this mass-loss 
process will eject enough mass to expose the high-temperature
core of the star, ionizing the gas inside the dust shell and 
forcing the transition from an AGB star to a planetary nebula.
Objects making this transition are in their ``post-AGB'' phase
(sometimes referred to as the ``proto planetary nebula'' phase).
Dust and gas characteristics change rapidly in this very short 
evolutionary stage, and the sources identified as post-AGB objects display 
a wide variety of properties. \citet{vw03} provides a good review of 
Galactic post-AGB objects and their general (albeit often disparate)
 properties. 
One of the challenges in understanding this phase of stellar evolution is 
identifying post-AGB candidates. Distinguishing post-AGB objects from other 
types of sources with circumstellar dust often depends on the derived 
luminosity for which a reliable distance is needed. While distances for 
field star candidates can be problematic, those for objects in the 
Large and Small Magellanic Clouds (LMC, SMC) are well constrained. 
\citet{wc01} identified 25 post-AGB candidates in the LMC using 8 \micron\
data from the {\em Midcourse Space Experiment} and near-IR $JK$ observations.
To date though, only two post-AGB objects have been identified in the SMC, 
IRAS 00350$-$7436 \citep{wfmc89} and [KVS2000] MIR 1 
\citep{kvs00}. Stellar evolution is also dependent on metallicity, but many
of the details are not well understood. 
Thus, identification and observation of additional post-AGB objects  in the 
LMC and SMC are therefore necessary for characterizing this phase of 
stellar evolution.

\section{Observations and Data Reduction\label{sec.obs}}

\begin{figure}
\plotone{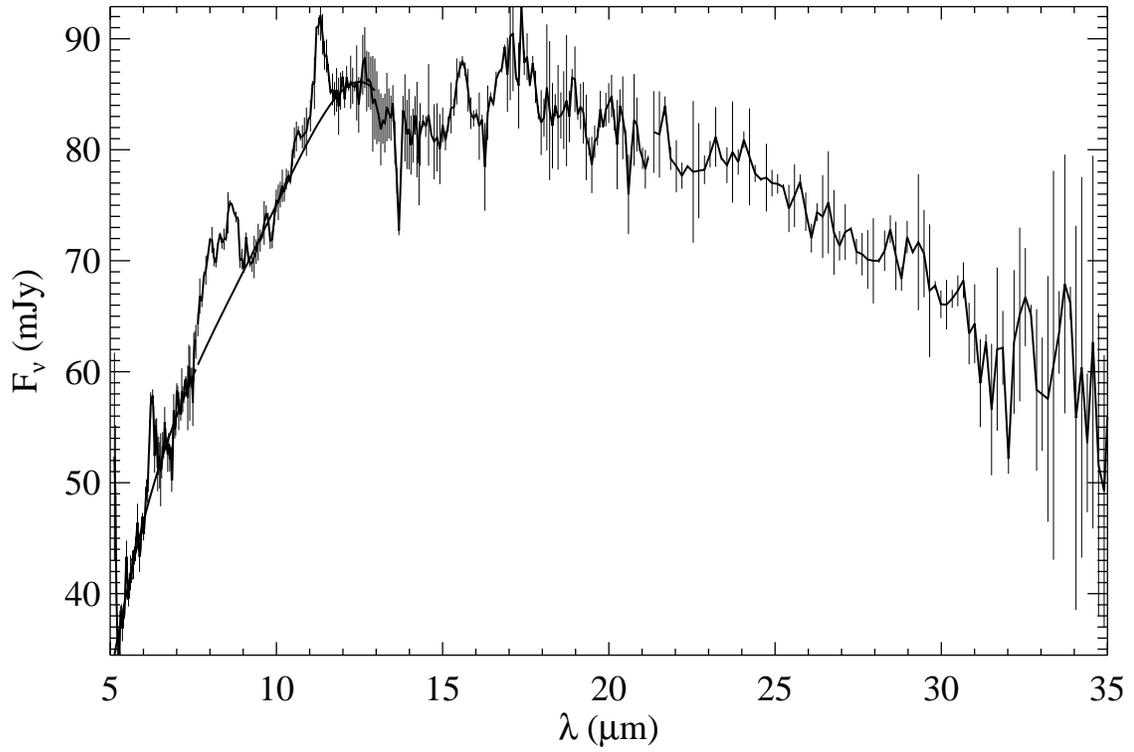}
%\plotone{figs/smc029_errorbars2_14b.eps}
\caption{The IRS spectrum of \ms. Uncertainties %(1 $\sigma$ error bars) 
reflect the differences between the two nods. The thin, smooth line from
5 \mum\ to 13 \mum\ shows
the continuum level subtracted from the spectrum to extract the PAH 
features (see text).
}\label{fig.smc029}
\end{figure}

We observed MSX SMC 029\footnote{We follow the nomenclature of 
\citet{evp01}. In the \msx\ Point Source 
Catalog V2.3 \citep{epk03}, MSX SMC 029 = G304.3649$-$43.5610.} with the 
Infrared Spectrograph \citep[IRS;][]{hou04} on the {\it Spitzer Space 
Telescope} \citep{wer04} on 2004 October 25 as part of a project to study 
circumstellar dust shells in the SMC.  The 
observations used  the Short-Low (SL) and Long-Low (LL) modules, which 
have a wavelength range of 5--36~\mum\ and an average spectral resolution of 
$\sim$100.  We extracted the spectra from the  \spi\ Science Center S13.2 (SL)
and S14.0 (LL) 
pipeline data.  Near-IR observations were made from 2004 November 25 to 
2005 September 19 at the Siding Spring Observatory (SSO) using the 
near-IR imaging system CASPIR.  \ms\ was detected only at 
H and K bands (2MASS also detected it, J00364631$-$7331351, at H and K 
but not at J).  The IRS and near-IR data reduction followed the standard 
procedures described in detail by \citet{skm06}. Table \ref{tab.phot} gives 
the near-IR photometry from both the SSO and 2MASS.  Figure \ref{fig.smc029} 
shows the IRS spectrum: MSX SMC 029 has a unique mid-IR spectrum, unlike any 
known Galactic or extragalactic object.

\section{Results  and Discussion\label{sec.results}}

\subsection{Evolutionary Status}

\ms\ shows the unusual combination of both \acet\ absorption at 13.7 \mum\ 
and polycyclic aromatic hydrocarbon (PAH) emission features. In its spectral 
energy distribution (SED), F$_\nu$ peaks at $\sim$ 17 \mum\ which corresponds 
to $T_d\approx280$ K, although the SED is quite broad so warmer and colder
components are also present. We estimated the IR luminosity L(1--100 \mum) 
$\sim4.5-5.5\times10^3$ \lsun\ by integrating the IRS spectrum and 
extrapolating longward of the IRS range with a 265 K graybody and shortward 
with a 530 K graybody, both
scaled to match the IRS data. In order to determine its likely evolutionary 
state, we examined Galactic spectra \citep{kspw02, skps03} from the Short 
Wavelength Spectrometer \citep[SWS, ][]{sws}  on the {\em Infrared Space 
Observatory} \citep{iso} and from IRS observations of LMC 
objects \citep{bkf06}. As noted above, no other object has this combination of 
mid-IR SED and spectral features: warm dust, PAH emission, and \acet\
absorption. We therefore discuss the possible evolutionary phases that \ms\ 
could represent.

\paragraph{Pre-Main Sequence?} Strong PAH emission is commonly observed in
H II regions and young stellar objects (YSOs), but \acet\ is not, 
particularly not at the strength observed in \ms.  Of the handful of YSOs 
in which weak \acet\ has been detected 
\citep[e.g.][and 
references therein]{lvd00}, only one has (very weak) PAH emission in its 
SWS spectrum. All have much colder SEDs than \ms, peaking longward of the
45 \mum\ SWS band edge, and have strong absorption 
from oxygen-rich silicates at 10 \mum\ and often CO$_2$ at 15.5 \mum. The
luminosity of \ms\ also much lower than the lowest luminosity H  
II region identified in the LMC sample of \citet{bkf06}, which have 
L(1--100 \mum) $>1.7\times  10^4 \lsun$. Most of the H II regions in the
LMC sample show evidence of being extended, such as flux jumps ~50\% at 14 
\micron\ (between the SL and LL modules) or extended emission in the IR
or visible images. \ms\ shows neither of these effects. It is unlikely, 
therefore, that \ms\ is either a YSO or an H II region.

\paragraph{Planetary Nebula?} PAHs have also been observed in Galactic 
planetary nebulae, but again \acet\ absorption has not. Young PNe typically 
have
dust temperatures of $\sim$150 K.  \ms\ does not have any emission
from the highly ionized lines often observed in PNe. While the lack of 
ionized lines  and warm dust do not strictly eliminate the PN possibility, 
the presence of \acet\ does.

\paragraph{AGB Star?} Another possibility is that \ms\ is an evolved star but
is still on the AGB. Carbon-rich AGB stars typically have \acet\ absorption
at 13.7 \mum, but they do not have PAH emission because they do not produce
sufficiently hard radiation fields to excite PAHs. Further, AGB stars
with similar near-IR colors ($H-K \gtrsim 2$) in the LMC have $K$-band 
amplitudes of $\Delta K \sim$ 1.5 mag \citep[e.g.][]{w98}, much greater than 
the $\Delta K$ = 0.43 mag observed in \ms\ (Table \ref{tab.phot}). The OGLE 
database \citep{ogle} contains a star $\sim$2\farcs2 from the 2MASS position 
with $I$ magnitude $\sim$19.2 but no variability above the noise. The 
presence of PAHs, together with the lack of strong $K$-band variability, 
implies that \ms\ cannot be an AGB star.

\paragraph{Post-AGB Object?} Only two spectra in the SWS database
show both PAH emission and \acet\ absorption at 13.7~\mum, IRAS
13416$-$6243 and AFGL 2688 (the Cygnus Egg), both post-AGB
sources.  We examined all of the spectra from carbon-rich
sources \citep[Groups 3, 4, and 5, subgroups CR, CT, CN, PU, U,
and UE defined by ][]{kspw02}.  On this basis alone, \ms\ looks
to be a post-AGB object.  No other possibility fits.  In addition,
as we show below, the PAH spectrum from \ms\ shows deviations
from the standard PAH spectrum consistent with those seen in other
post-AGB objects.  AFGL 2688 is thought to be only
a few hundred years past the AGB stage \citep{hgpc02}.  The dust
around \ms\ is similar, significantly cooler than typically seen
around AGB stars, so it is probably of a similar age.  Because
the dust around \ms\ is warmer than that around AFGL 2688, it may
 be even closer to the transition.

\subsection{Solid State Features}

\begin{figure}
\plotone{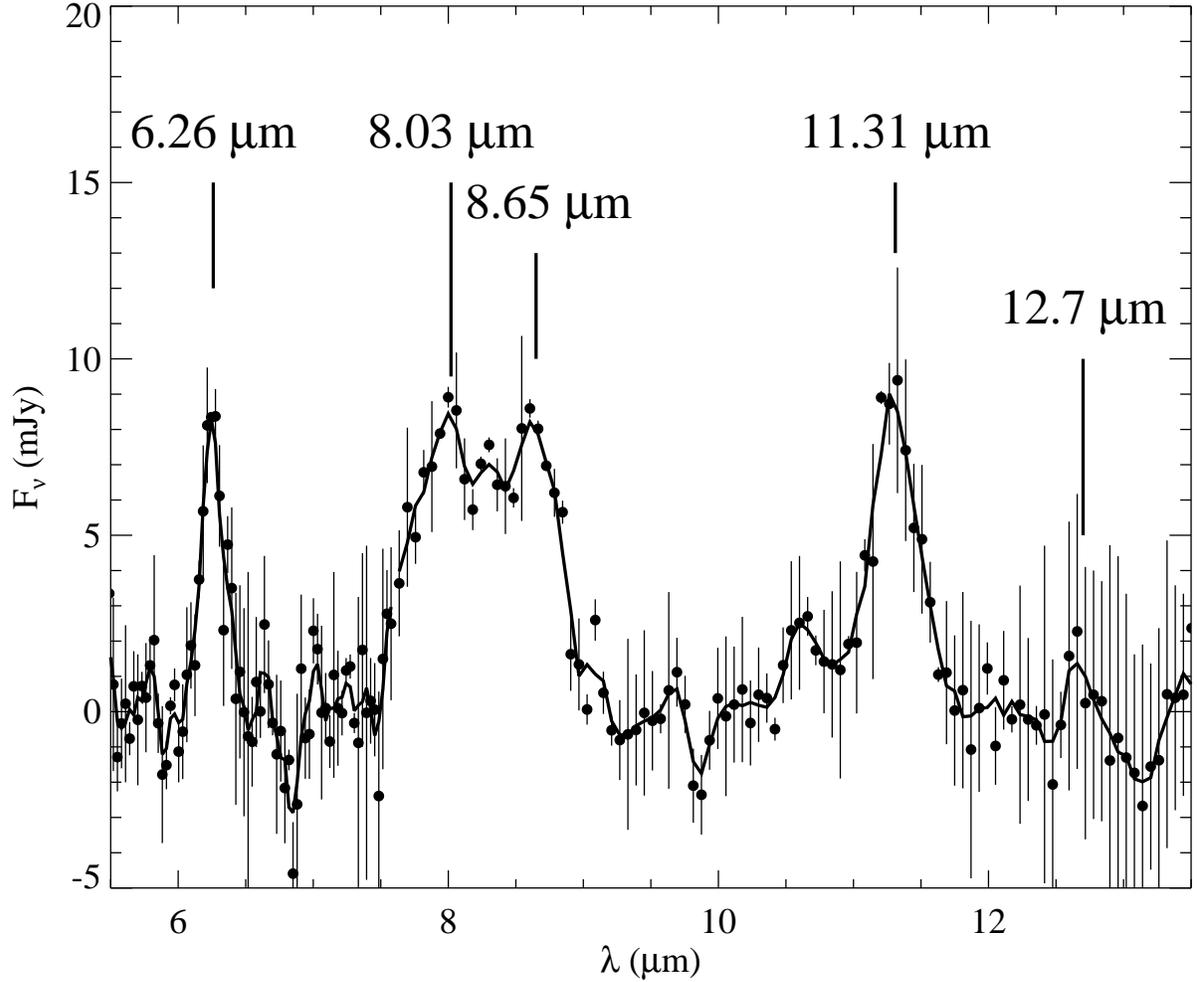}
%\plotone{figs/pah029_14b.ps}
\caption{The continuum-subtracted PAH spectrum for \ms. The data are the filled
circles with error bars. The line shows the spectrum after 
smoothing with a 3-pixel boxcar. The labels indicate the central 
wavelengths of the PAH features. The 12.7 \mum\ feature was only tentatively detected. 
}
\label{fig.pahs}
\end{figure}

To extract the PAH features for analysis, we fit a 6$^{th}$-order
polynomial to the spectrum between 5 and 13 \mum\ (smooth line in
Fig. \ref{fig.smc029}). Figure \ref{fig.pahs} shows the
`continuum'-subtracted result,  and Table \ref{tab.pahs}
contains the central wavelengths and fluxes of the detected
features. Following \citet{skf05}, we extracted the PAH strengths
by trapezoidal integration of the emission above a line
fit to the continuum to either side using the wavelength ranges
given in Table \ref{tab.pahs}.  The central wavelength $\lambda_c$
is the wavelength at which half the integrated flux
lies to the red and half to the blue. Table \ref{tab.pahs} also
includes fluxes for the two components of the 7--9 \mum\ feature
complex.  The flux at 12.7 \mum\ is an upper limit;
higher-sensitivity data are needed to confirm this possible feature. 
Confident detection of a PAH feature at $\sim$ 17 \mum\ is precluded
by the molecular absorption in the region (see \S \ref{sec.abs}).

\begin{figure}
\includegraphics[width=5in]{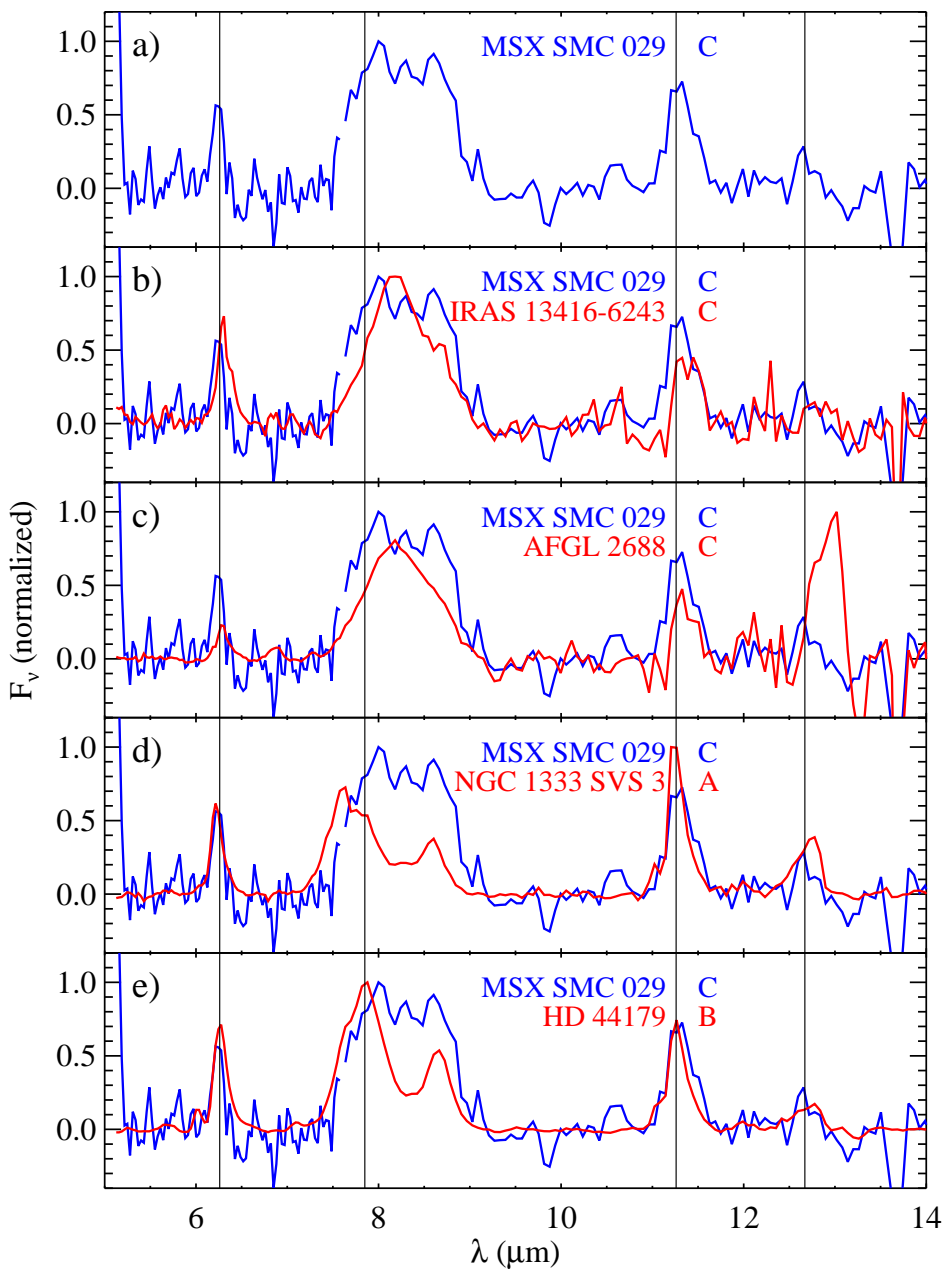}
%\plotone{f3.eps}
%\plotone{figs/f3_14b.eps}
\caption{ The 7--9 \mum\ complexes from \ms\ (blue line) compared
with (red lines) (b) the post-AGB object IRAS 13416$-$6243 (class C PAHs); 
(c) the post-AGB object AFGL 2688 (class C); (d) the young stellar object
NGC 1333 SVS 3 (class A); and (e) the planetary nebula HD 44179 (class B).
All the spectra have had the continuum approximated with a polynomial and
subtracted. The spectral resolution of the SWS data  was degraded to match
that of IRS. Vertical lines are at the typical class B positions of 6.26, 7.85,
11.2, and 12.7 \mum.
}\label{fig.8comp}
\end{figure}

The most striking characteristic of the PAH features is their
wavelength shift compared to more typical PAH spectra, as shown
in Figure \ref{fig.8comp}.  Adopting
the classification introduced by \cite{phv02}, the PAHs in \ms\
are class C, a category with only two sources from the SWS
database, IRAS 13416 and AFGL 2688, and more recently, HD 233517,
an O-rich AGB star with PAH emission from a circumstellar disk
 observed by the IRS \citep{jbs06}.  In those spectra,
the 6.2~\mum\ feature is shifted to 6.3~\mum, and the peak of the
7--9~\mum\ PAH complex, which usually occurs at 7.65 or 7.85~\mum,
is shifted to the red of 8.0~\mum.  In \ms, these features are
centered at 6.26 and 8.23~\mum, respectively.  Excluding the
8.6~\mum\ C--H in-plane bending mode by fitting a line under it
shifts the center from 8.23 to 8.12~\mum.  Alternatively, fitting
Gaussians to the two components produces centers at 8.03 and
8.65~\mum.  However one measures it, the  C--C modes usually
seen at 7.65 and 7.85~\mum\ here appear redward of 8.0~\mum.

The positions and relative strengths of the PAH features have been attributed 
to the relative dominance of the ionization fraction of the 
PAHs, their composition, their isotopic ratios, their size, and the degree of 
UV processing they 
have undergone, to name a few. \cite{vpt02}, for example, examined the role 
of several of these factors in a small sample of LMC HII region spectra and 
suggested the molecular structure, which influences which C$-$C and C$-$H modes
are present, may control the strengths in the LMC sources.
\cite{bt05} found that the relative strengths of the 7.65 and
7.85~\mum\ features in extended reflection nebulae vary, with the
dominant emission shifting from 7.65 to 7.85~\mum\ as the UV field
grows weaker.  \cite{skf05} interpreted their PAH spectra from
Herbig AeBe (HAeBe) stars similarly.  These spectra are excited by
weak UV fields, and the peak component in this region has shifted
further to the red, to 7.9--8.0~\mum.  The spectrum of \ms\ follows
this trend, with the centroid of the C$-$C modes at 8.1--8.2~\mum. 

Other characteristics of the PAH spectrum from \ms\ are consistent
with this shift in the C$-$C modes.  The entire 7--9~\mum\ PAH
complex is only $\sim$4 times stronger than the out-of-plane
C$-$H bending mode at 11.3~\mum, compared to 8--30 times stronger
in the HAeBe stars \citep{skf05}, indicating that the ionization
fraction of the PAHs is lower in \ms.  \cite{hvp01} studied the
PAH features in the 11--13~\mum\ region in 16 SWS spectra and found that
the relative strengths of the solo and trio C$-$H out-of-plane
bending modes at 11.3 and 12.7~\mum\ varied with ionization
fraction, with more ionized PAHs having a stronger trio mode at
12.7~\mum.  They interpreted this result as an indication of PAH
processing, with more unprocessed PAHs having long, regular edges
and thus more emission in the solo mode.  In \ms, the ratio of
the 12.7~\mum\ trio mode to the 11.3~\mum\ solo mode is at most
0.25 (using the upper limit at 12.7~\mum), as low as any in the
sample examined by \cite{hvp01}.  They also show that this ratio
decreases as the ionization fraction decreases.  The PAH spectrum
of \ms\ follows this trend, indicating both a low ionization
fraction (low 7--9~\mum/11.3~\mum\ ratio) and relatively
unprocessed PAHs (low 12.7~\mum/11.3~\mum\ ratio).

\subsection{Molecular Absorption Features\label{sec.abs}}

\begin{figure}
\plotone{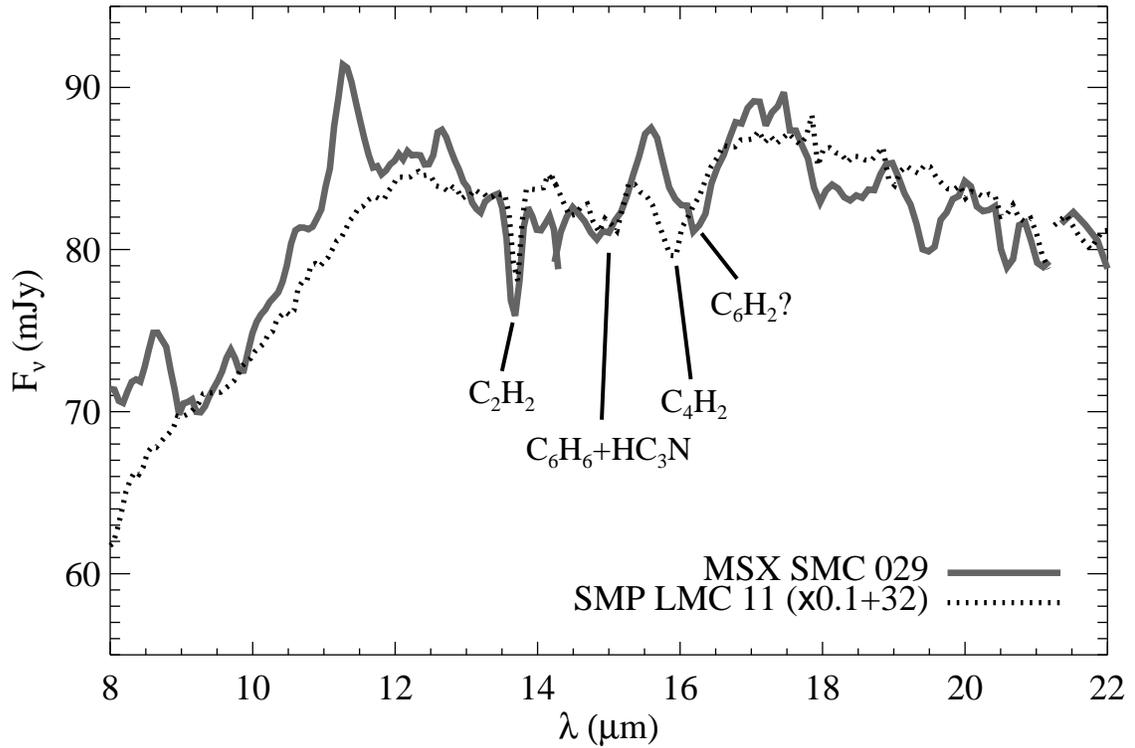}
%\plotone{figs/s011compb_14b.ps}
\caption{Comparison of \ms\ (solid line) with the IRS spectrum 
from the post-AGB object SMP LMC 11 (dotted line, scaled and offset to 
facilitate 
comparison). The molecular absorption features identified in the high
resolution spectrum of SMP LMC 11 \citep{bss06} and AFGL 2688 \citep{cht01}
are
indicated.
}\label{fig.011}
\end{figure}

\citet{cht01} recently made the first detections
of the polyacetylenic molecules \diac\ (diacetylene) and C$_6$H$_2$ outside 
the Solar System in the very young  post-AGB objects AFGL 618 and AFGL 2688, 
as well as detecting 
the more familiar bands of \acet\ (acetylene) at  13.7 \mum\ and HCN at 14 
\mum. In \ms, the \acet\ band  is the most prominent and easily 
identified molecular absorption feature. The limited signal-to-noise 
ratio and trouble determining a continuum level make extracting other features 
challenging. Fortunately, \citet{bss06} recently 
observed the post-AGB object SMP LMC 11 with both the high- and low-resolution 
modules of  IRS. They identified several hydrocarbons in the 
high-resolution spectrum whose fingerprints are also clearly discernible 
in the low-resolution spectrum. In addition to \acet, their spectrum shows
absorption from \ben\ and \cyac\ (benzene and cyanoacetylene) at 
$\sim$15 \mum\ 
and \diac\  at $\sim$16 \mum.  Figure \ref{fig.011} compares
\ms\ to SMP LMC 11. The similarity of the spectra leads us to suggest that 
the same molecules may be shaping the \ms\ spectrum. The absorption
near 16 \mum\ in \ms\ has an additional red component which could be from 
C$_6$H$_2$ as was seen in AFGL 618 \citep{cht01}. Absorption from 
the P and R branches of the \acet\ feature may affect the appearance
of the spectra as well. The strong mid-IR absorption and non-detection
at $J$ band, combined with a preliminary model that suggests  high
visual opacity, imply the presence of very dense material. This could
a residual circumstellar envelope from the AGB phase, which would be 
consistent with the youth of the post-AGB object, although such
a nebula may require more extreme youth than is likely, or the material could
be in the form of a dense disk.

\section{Summary \label{sec.conc}}

We obtained a 5--35 \mum\ low-resolution spectrum from  \ms\ with the 
\spi\ IRS. It has an unusual
spectral energy distribution and combination of mid-IR spectral 
features. \ms, like the post-AGB objects AFGL 2688 
and IRAS 13416$-$6243, shows both \acet\ absorption 
at  13.7 \mum\ and PAH emission features. The PAHs in \ms\ are in 
the rare C class of \cite{phv02}. The shape of the
7--9 \mum\ PAH complex and the relative flux of the
11.3 \mum\ feature to that of the 12.7 \mum\ feature suggest that \ms\
has a low ionization fraction or unprocessed PAHs. The 13--16 \mum\ wavelength
region is similar to that of SMP LMC 11 and may be shaped by the
same absorption features that have been resolved in the LMC source and in 
AFGL 618: 
\acet, \diac, \cyac, \ben, and possibly C$_6$H$_2$. 
We suggest that \ms\ is a very young
post-AGB object, only the third known in the SMC.

\acknowledgements
We thank M. Matsuura for useful discussions on molecular absorption
and color-color diagrams and the referee for helpful suggestions 
that improved the paper.
This work is based on observations made with the 
{\em Spitzer Space Telescope} which is operated by JPL/Caltech
 under NASA contract 1407. Support for this work was 
provided in part by NASA through contract
number 1257184 (G. C. S. and J. B.-S.); P. R. W. received funding for 
this work from the 
Australian Research Council. This research has made use of NASA's 
Astrophysics Data System Bibliographic Services, data products from 2MASS
 which is a joint project of the University of Massachusetts and IPAC/Caltech
funded by NASA and the NSF,  and the Simbad database 
operated at CDS, Strasbourg, France.

\clearpage

\begin{deluxetable}{cccl} 
\tablewidth{0pt}
\tablecaption{Near-IR Photometry of MSX SMC 029 \label{tab.phot}}
\tablehead{ \colhead{Date} & \colhead{H} & \colhead{K} & 
\colhead{Source} }
\startdata
980808 & 15.75$\pm$0.18 & 13.50$\pm$0.04 & 2MASS \\
041125 & 15.79$\pm$0.07 & 13.75$\pm$0.17 & SSO   \\
050126 & \nodata       & 13.67$\pm$0.06 & SSO   \\
050312 & \nodata       & 13.92$\pm$0.08 & SSO   \\
050724 & \nodata       & 13.74$\pm$0.03 & SSO   \\
050919 & \nodata       & 13.74$\pm$0.05 & SSO   
\enddata
\tablecomments{Dates are in the form YYMMDD. {\em K}-band wavelengths are  
{\em K$_s$}=2.17 \mum\ for 2MASS and {\em K}=2.22 \mum\ for SSO} 
\end{deluxetable}

\clearpage

\begin{deluxetable}{cccc} % Table 1
\tablewidth{0pt}
\tablecaption{PAH Wavelengths and Fluxes\label{tab.pahs}}
\tablehead{
\colhead{$\lambda_c$} & \colhead{Flux} &
\colhead{$\lambda_{blue}$} & \colhead{$\lambda_{red}$}\\
\colhead{(\mum)} & \colhead{($ 10^{-16}$ W m$^{-2}$)}&
\colhead{(\mum)} & \colhead{(\mum)} 
}
\startdata
 6.26$\pm$0.03   &   $1.55\pm0.27$   &  5.80--5.95  & 6.65--6.80   \\
 8.23$\pm$0.06   &   $3.38\pm0.24$   &  7.30--7.60  & 9.00--9.30   \\
11.32$\pm$0.05   &  $0.64\pm0.10$    & 10.80--11.05 & 11.85--11.95 \\
12.69              &  $<0.15$            & 12.35--12.60 & 13.25--13.40 \\
8.03               & 2.50                & \nodata      & \nodata \\
8.65               & 1.41                & \nodata      & \nodata
\enddata
\tablecomments{$\lambda_{blue}$ and $\lambda_{red}$ are the 
ranges over which the line segments were fit for the flux integration.
Approximate fluxes for the components of the 7--9 \mum\ complex
are from integrating the Gaussian fits. Thus, the sum of the fluxes
does not equal the total given for 8.25 \mum.}
\end{deluxetable}

\clearpage

\clearpage

\clearpage

\clearpage

\end{document}